\documentclass[conference]{IEEEtran}
\IEEEoverridecommandlockouts
\usepackage{cite}
\usepackage{amsmath,amssymb,amsfonts}
\usepackage{algorithmic}
\usepackage{graphicx}
\usepackage{textcomp}
\usepackage{xcolor}
\usepackage{pdfpages}
\def\BibTeX{{\rm B\kern-.05em{\sc i\kern-.025em b}\kern-.08em
    T\kern-.1667em\lower.7ex\hbox{E}\kern-.125emX}}

\usepackage{paralist}

\usepackage[hyphens]{url}
\usepackage{hyperref}
\usepackage{booktabs}

\usepackage{float}
\usepackage{csquotes}


\usepackage{xpatch}
\xpatchcmd\IEEEkeywords{---}{-}{}{}

\makeatletter
\renewcommand{\fnum@figure}{Figure~\thefigure}
\makeatother

\begin{document}

\title{\bfseries\Large Does Johnny Get the Message? \\ Evaluating Cybersecurity Notifications for Everyday Users}

\author{
\IEEEauthorblockN{Victor Jüttner}
\IEEEauthorblockA{
Dept. of Computer Science, Leipzig University\\
Center for Scalable Data Analytics and Artificial\\
Intelligence (ScaDS.AI) Dresden/Leipzig, Germany \\
e-mail: victor.juettner@cs.uni-leipzig.de}
\and
\IEEEauthorblockN{Erik Buchmann}
\IEEEauthorblockA{
Dept. of Computer Science, Leipzig University\\
Center for Scalable Data Analytics and Artificial\\
Intelligence (ScaDS.AI) Dresden/Leipzig, Germany \\
e-mail: erik.buchmann@cs.uni-leipzig.de}
}

\maketitle

\begin{abstract}
Due to the increasing presence of networked devices in everyday life, not only cybersecurity specialists but also end users benefit from security applications such as firewalls, vulnerability scanners, and intrusion detection systems. Recent approaches use large language models (LLMs) to rewrite brief, technical security alerts into intuitive language and suggest actionable measures, helping everyday users understand and respond appropriately to security risks. However, it remains an open question how well such alerts are explained to users. LLM outputs can also be hallucinated, inconsistent, or misleading. In this work, we introduce the Human-Centered Security Alert Evaluation Framework (HCSAEF). HCSAEF assesses LLM-generated cybersecurity notifications to support researchers who want to compare notifications generated for everyday users, improve them, or analyze the capabilities of different LLMs in explaining cybersecurity issues. We demonstrate HCSAEF through three use cases, which allow us to quantify the impact of prompt design, model selection, and output consistency. Our findings indicate that HCSAEF effectively differentiates generated notifications along dimensions such as intuitiveness, urgency, and correctness.
\end{abstract}



\section{Introduction}
\label{sec:intro}

To ward off cyberattacks, security applications such as firewalls~\cite{ingham2002history, Liang2022firewallevolution}, vulnerability scanners~\cite{tundis2018vulnscanner}, or intrusion detection systems (IDS)~\cite{patel2010survey} scan networks and/or connected devices and generate security alerts about suspicious activity.
For example, an IDS might identify unusual network packets and report:
\enquote{\textit{HTTP Response abnormal chunked for transfer encoding}}.
A firewall might log the alert:
\enquote{\textit{Wsmprovhost.exe trying to connect to 203.0.113.25:443, Connect Layer, Layer Run-Time ID 48}}.
A vulnerability scanner may produce:
\enquote{\textit{Remote Desktop Protocol RCE Vulnerabilities (2671387) detected. CVSSv3 Score 9.7. CVE-2012-0002 CVE-2012-0152 DFN-CERT-2012-0477}}.
Such alerts typically require expert interpretation, must be analyzed in the context of the network setup, and translated into meaningful countermeasures if necessary.

Because of the widespread proliferation of smart, connected devices, everyday users without cybersecurity expertise are increasingly required to protect complex networks and could benefit from such security applications.
Recent work~\cite{juettner2024chatids, hoffmann2024ChatSEC} uses large language models (LLMs) to rewrite cybersecurity alerts into intuitive notifications (see Figure~\ref{fig:gpt4oresponse}). These notifications aim to explain the nature of the security threat and suggest actionable countermeasures.
However, it is challenging to assess whether the LLM-generated notifications actually provide helpful advice. LLMs can generate superficial notifications that fail to address specific threats. They may substitute one unintuitive technical term for another, hallucinate, or produce inconsistencies. LLMs may also provide incorrect or unsafe advice. Even slight changes in the model or prompt can result in significantly different notifications.

Thus, researchers need to conduct multi-faceted analyses, compare LLMs based on their ability to rewrite cybersecurity alerts into self-explanatory notifications, optimize prompting strategies, and examine the robustness of the generated notifications. Our research question is as follows:

\textbf{How can we systematically evaluate the quality of generated cybersecurity notifications?}

We propose the \textit{Human-Centered Security Alert Evaluation Framework (HCSAEF)} to assess the wording of cybersecurity alerts across seven dimensions:
\textit{Consequences},
\textit{Context},
\textit{Countermeasures},
\textit{Correctness},
\textit{Intuitiveness},
\textit{Personalization}, and
\textit{Urgency}.
These dimensions are derived from existing work on general LLM evaluation frameworks~\cite{chang2024llmEvalStudy, mehri-eskenazi-2020-unsupervised} and research in security communication~\cite{rogers1983cognitive, bravo-lillo, wogalter2018communication, bartsch2013contextualized, Kauer2012ItIN, warning-eeg, zimmermann2024beyondFear, sasse}.
We make three contributions:
\begin{compactitem} \item We devise HCSAEF, our Human-Centered Security Alert Evaluation Framework for cybersecurity alerts.
\item We define HCSAEF’s seven evaluation dimensions using a 5-point Likert scale.
\item We demonstrate the applicability of HCSAEF with a case study in which an LLM rewrites IDS alerts.
\end{compactitem}

Our study acknowledges that HCSAEF is sensitive to prompt design, reflecting trade-offs between contextual richness and output precision, especially regarding personalization and actionability. We identify key differences between LLMs, e.g., Grok 3’s strong personalization versus DeepSeek R1’s unsafe advice and capture consistency trends across multiple repetitions. These insights help guide model selection, prompt refinement, and the design of effective LLM-generated security notifications. Currently, HCSAEF is applied manually. Next, we will automate it with an LLM-as-a-Judge approach~\cite{Zheng2023LLMasAJudge}.

\textit{Paper outline:} Section~\ref{sec:related} reviews related work. Section~\ref{sec:approach} describes the HCSAEF approach, followed by a case study in Section~\ref{sec:case}. Section~\ref{sec:conclusion} concludes the paper and discusses future work.

\newcommand{\myspc}[0]{\\[-1.5mm]}
\begin{figure*}[htb]
    \centering
\fbox{\parbox{.92\linewidth}{\small
\textbf{\large Urgent Security Alert – Suspicious Network Activity Detected in Your Home Network}\myspc

Dear [Homeowner's Name],\myspc

I hope this message finds you well. I am reaching out to inform you about a security alert detected on your home network. The network monitoring tool has flagged an issue with a message labeled:\myspc

\textbf{\em \enquote{HTTP Response abnormal chunked for transfer-encoding.}}\myspc

This means that unusual activity has been detected in the way data is being transferred over your network. While this does not immediately confirm an attack, it is a strong indication that one of your smart home devices or computers might be compromised or targeted by hackers.\myspc

\textbf{What This Means for You}\myspc

    This alert suggests that a device on your home network is sending or receiving data in an abnormal way.
    It could indicate a hacker attempting to gain access to your smart home devices (e.g., security cameras, smart thermostats, or even your personal computer).
    If this is an attack, the hacker may be trying to steal personal information, spy through a connected camera, or gain control over smart devices.\myspc

\textbf{How Urgent Is This?}\myspc

This is a serious security concern that should be addressed immediately.
If left unchecked, your devices and personal data may be at risk. Cybercriminals can potentially access sensitive information, misuse smart home devices, or even use your network for illegal activities.\myspc

\textbf{What You Should Do Now – Step-by-Step Guide to Secure Your Network}\myspc

Follow these steps to protect your home network and devices from potential harm:\myspc

    \textit{1. Restart Your Internet Router:}
        Unplug your router from the power source, wait 30 seconds, then plug it back in.
        This may stop an ongoing attack or suspicious activity.\myspc

    \textit{2. Identify Devices on Your Network:}
        Use your router’s app or web interface to check all connected devices.
        If you see any unfamiliar devices, disconnect them immediately.\myspc

    \textit{3. Update Your Smart Devices and Computers:}
        Go to each device’s settings and check for software updates.
        This includes smart cameras, thermostats, TVs, laptops, and phones.
        Keeping devices updated prevents hackers from exploiting vulnerabilities.\myspc

    \textit{4. Change Your Wi-Fi Password:}
        Use a strong password with at least 12 characters, including letters, numbers, and symbols.
        Reconnect your devices using the new password.\myspc

    \textit{5. Check Your Security Cameras (If Installed):}
        If you have smart security cameras, review recent recordings to see if they have been accessed unexpectedly.
        Change their passwords to something strong and unique.\myspc

    \textit{6. Scan Your Computer for Malware:}
        If you use a Windows or Mac computer, run a full antivirus scan to check for infections.\myspc

    \textit{7. Monitor for Further Alerts:}
        If you continue receiving alerts about suspicious network activity, you may need professional help to secure your network.\myspc

\textbf{Final Thoughts}\myspc

If you follow these steps, your home network will be significantly safer. However, if the suspicious activity continues, it may indicate a deeper security issue requiring expert intervention.

I highly recommend taking these steps immediately to ensure your privacy and safety. If you need any help or have questions, please feel free to reach out.\myspc

Stay safe,
[Your Name]
Cybersecurity Specialist

}}
    \caption{Example of a cybersecurity alert rewritten by GPT-4o into a detailed, user-friendly notification tailored for non-expert homeowners.}  
    \label{fig:gpt4oresponse}
\end{figure*}

\section{Related Work}
\label{sec:related}


\subsection{Smart Home Threat Landscape and Intrusion Detection}

Modern smart homes are equipped with a variety of interconnected devices—ranging from smart TVs and refrigerators to thermostats and lighting systems—that enhance convenience and automation. However, these devices often suffer from inadequate security measures, such as the lack of regular firmware updates, making them attractive targets for cyberattacks~\cite{smart-device-vulns}. Their interconnected nature means that a compromise in one device can potentially lead to a breach across the entire home network~\cite{smart-network-issues}. This risk is further amplified by the fact that many users lack the technical expertise needed to properly configure and secure these devices~\cite{Pattnaik2022ASO}.

To mitigate these risks, considerable research has been directed toward the development of IDS tailored for smart home environments. Anthi et al.~\cite{AnthiIDS2018} introduced a supervised IDS capable of detecting various network-based attacks in IoT environments. Sikder et al.~\cite{aegis} developed Aegis+, a context-aware and platform-independent security framework that provides users with detailed, customizable alerts about malicious activity, including the type of event, affected devices, and their physical locations. Similarly, the Dynamic Risk Assessment Framework (DRAF) proposed by Collen and Nijdam~\cite{draf} dynamically assesses IoT threats and adjusts alerts based on user-defined risk thresholds. Visoottiviseth et al.~\cite{piti} presented PITI, a hybrid IDS that enhances user awareness by delivering auditory and textual alerts with detailed information about detected attacks and the IP addresses of affected devices.


\subsection{Usable Security Notifications}

Security alerts aim to warn users before harm occurs, but their effectiveness often suffers due to misunderstandings, lack of trust, or perceived inconvenience, especially among non-experts~\cite{purposeofwarnings, jones2022non}. Fear-based messaging, while tempting, has proven ineffective and can erode trust~\cite{sasse, depuis}.

Instead, effective alerts should use brief, nontechnical language~\cite{bravo-lillo, bauer}, clearly explain the risk~\cite{bravo-lillo}, the consequences of ignoring it~\cite{bravo-lillo}, and how the threat could personally affect the user~\cite{bartsch2013contextualized, Kauer2012ItIN}. Alerts should also provide actionable steps for mitigation~\cite{bravo-lillo}, ideally in a way that aligns with users' mental models~\cite{bartsch2013contextualized}.

Theories such as Protection Motivation Theory (PMT)~\cite{rogers1983cognitive} and the Communication-Human Information Processing (C-HIP) model~\cite{wogalter2018communication} support this approach by emphasizing the roles of perceived severity, response efficacy, and cognitive processing in user behavior. Cranor~\cite{cranor-human-in-loop} and Zimmermann et al.~\cite{ZIMMERMANN2019169} further advocate for human-centered security, shifting the focus from human error to system support.



\subsection{LLMs for Cybersecurity Communication}


LLMs increasingly influence many aspects of cybersecurity~\cite{hang2025llm-cybersec-review}, one of which is their ability to translate technical outputs—such as IDS alerts and vulnerability reports—into formats understandable by non-experts.

ChatIDS~\cite{juettner2024chatids}, introduced by Jüttner et al., utilizes GPT-3.5-turbo to translate IDS alerts into user-friendly security notifications tailored for non-expert users in smart home environments. Similarly, ChatSEC~\cite{hoffmann2024ChatSEC}, developed by Hoffmann and Buchmann, employs GPT-4 to transform vulnerability scan results into accessible explanations, supporting university network administrators with limited IT security expertise. HuntGPT~\cite{ali2023huntgpt}, introduced by Ali and Kostakos, combines machine learning-based IDS with explainable AI and GPT-3.5-turbo to provide analysts with actionable threat explanations through a conversational dashboard. SHIELD~\cite{gandhi2025shield}, proposed by Gandhi et al., integrates statistical anomaly detection, graph-based analysis, and LLM reasoning to detect and explain advanced persistent threats, offering interpretable attack narratives to security analysts.


\subsection{Prompt Strategies}

Prompt engineering is the practice of designing inputs to LLMs to improve the accuracy and relevance of their outputs. How a task is framed through role assignment, structured instructions, or contextual information can strongly influence model behavior. Common strategies include chain-of-thought prompting, self-reflection, and persona conditioning. For example, assigning the model the role of an expert or breaking down a complex instruction into steps can lead to more coherent and useful responses. These techniques help align model inference with user intent, especially in domains that require clarity for non-expert users~\cite{schulhoff2024prompt}.

Current state-of-the-art models include DeepSeek R1~\cite{deepseek2024}, OpenAI's GPT-4o and O1~\cite{openai2024gpt4o, openai2023gpto1}, and Grok 3 from xAI\cite{xai2024grok}. While each model varies in architecture and behavior, their performance is strong in natural language reasoning, code generation, and multimodal inference, according to multiple benchmarks~\cite{llmstats, lmarena}.

\subsection{Qualitative Evaluation of LLM Responses}

Automated reference-based metrics like BERTScore~\cite{zhang2020bertscore} and MoverScore~\cite{zhao-etal-2019-moverscore} fall short when applied to open-ended language tasks, where valid responses can vary widely in form. Their limitations in capturing semantic nuance or conversational appropriateness have been well documented~\cite{liu-etal-2016-evaluate}, motivating a shift toward qualitative evaluation strategies.

To automate evaluation frameworks such as OpenAI Evals~\cite{openai_evals} and G-Eval~\cite{liu2023geval} have emerged. OpenAI Evals provides a structured environment for benchmarking across diverse tasks, while G-Eval uses LLMs as evaluators to assess dimensions like correctness, coherence, and helpfulness.

Recent work has further refined the dimensions used in qualitative evaluation. Chang et al.\cite{chang2024llmEvalStudy} identify key criteria such as factual accuracy, relevance to the prompt, fluency, transparency in reasoning, safety in terms of avoiding harmful or misleading content, and general alignment with human values. In a conversational context, the FED framework\cite{mehri-eskenazi-2020-unsupervised} introduces similar but dialogue-specific dimensions, focusing on contextual relevance, logical coherence, natural phrasing, factual correctness, and user engagement.

In domain-specific settings like cybersecurity, the SECURE benchmark~\cite{bhusal2024secure} evaluates LLMs on tasks that require contextual understanding, factual consistency, and reasoning over real-world advisories. Its focus on practical, high-stakes scenarios makes it a valuable reference for qualitative evaluation in specialized domains.

\section{Our HCSAEF approach}
\label{sec:approach}

We introduce HCSAEF, our Human-Centered Security Alert Evaluation Framework, to evaluate LLM-generated cybersecurity notifications across seven dimensions.
We adapted the dimensions \textit{Context}, \textit{Correctness}, and \textit{Intuitiveness} from general LLM evaluation frameworks~\cite{chang2024llmEvalStudy, mehri-eskenazi-2020-unsupervised}, which focus on accuracy, relevance, and clarity.
The remaining dimensions, \textit{Countermeasures}, \textit{Consequences}, \textit{Personalization}, and \textit{Urgency}, were derived from security communication research. In particular, \textit{Countermeasures} and \textit{Consequences} reflect Protection Motivation Theory and the need for actionable, motivating content~\cite{rogers1983cognitive, bravo-lillo, wogalter2018communication}. \textit{Personalization} improves relevance to the user~\cite{bartsch2013contextualized, Kauer2012ItIN}, while \textit{Urgency} emphasizes timely action without relying on fear appeals~\cite{warning-eeg, zimmermann2024beyondFear, sasse}.

We rate each dimension on a 5-point Likert scale from 0 to 4, which aligns with common practice in this field.
The lowest rating, 0 (\textit{Unsatisfactory}), means that this dimension is not present in the notification.
1 (\textit{Needs Improvement}) suggests that the dimension is present but not adequately worded.
2 (\textit{Satisfactory}) refers to a clearly identifiable dimension.
3 (\textit{Very Good}) indicates a dimension that is well fulfilled.
Finally, 4 (\textit{Outstanding}) means that the dimension exceeds expectations.
In the following, we explain each dimension in alphabetical order and describe how it is rated.

\paragraph{Consequences}  
The dimension \textbf{Consequences} (see Table~\ref{tab:dim:consequences}) measures whether the consequences of disregarding the particular alert are communicated to the user.

\begin{table}[h]
    \caption{Definition of the dimension \enquote{Consequences}.}
    \label{tab:dim:consequences}
    \centering
    \begin{tabular}{@{}c p{0.8\columnwidth}@{}}
        \toprule
        \textbf{\em Scale} & \textbf{\em Definition} \\
        \midrule
        0 & The notification does not mention consequences. \\
        1 & The consequences are mentioned at a superficial level, e.g., \enquote{Not acting could result in a loss of data.} \\
        2 & General consequences are mentioned without details, e.g., \enquote{Someone could steal personal data from your devices.} \\
        3 & Specific consequences for the home network are mentioned, e.g., \enquote{This could lead to data theft, financial or legal problems, or even your smart home devices being used for espionage.} \\
        4 & The notification names specific consequences along with the affected devices, e.g., \enquote{An attacker could eavesdrop on your conversations with your Echo Hub or track movement with your Shelly Motion Sensor.} \\
        \bottomrule
    \end{tabular}
\end{table}

For example, the consequences of disregarding a successful denial-of-service attack on a smart device are typically low. The user could simply wait out the attack until the device is working again.  
Non-existent, superficial, or generic consequences result in lower ratings. What a user without cybersecurity expertise actually needs is an explanation of the consequences that is specific to their network setup or, even better, specific to their network and the devices present on it.

\paragraph{Context}  
Dimension \textbf{Context} (see Table~\ref{tab:dim:context}) reflects how well the cybersecurity threat is explained.  
The user needs this information to understand what the threat means for the security of their home.

\begin{table}[h]
    \caption{Definition of the dimension \enquote{Context}.}
    \label{tab:dim:context}
    \centering
    \begin{tabular}{@{}c p{0.8\columnwidth}@{}}
        \toprule
        \textbf{\em Scale} & \textbf{\em Definition} \\
        \midrule
        0 & The notification does not mention the context of the threat. \\
        1 & The context is mentioned at a superficial level, e.g., \enquote{Malicious software, designed to damage or disrupt systems, could steal data or gain unauthorized access.} \\
        2 & General contextual information is provided, e.g., \enquote{There is traffic inside your network that looks as if it is related to a type of malware called the Harakit botnet.} \\
        3 & Specific context about the attack mechanism is given, e.g., \enquote{Imagine your router as a locked door, and a hacker trying to trick the lock and enter your network uninvited.} \\
        4 & Detailed information about all concepts needed to understand the cybersecurity threat without reading external sources. \\
        \bottomrule
    \end{tabular}
\end{table}

For example, it is important to understand whether a threat is about reconnaissance and preparation for an attack, or an ongoing attack. The scale for this dimension ranges from not mentioning the context (0) to explaining the threat in great detail (4), so that the user does not need external information sources to fully understand the threat.

\paragraph{Countermeasures}

Dimension \textbf{Countermeasures} (see Table~\ref{tab:dim:countermeasures}) is about explaining countermeasures that are appropriate to ward off the cybersecurity threat. A countermeasure is satisfactory if it is rather broad and unspecific but generally applicable and mitigates the threat to some extent.

\begin{table}[h]
    \caption{Definition of the dimension \enquote{Countermeasures}.}
    \label{tab:dim:countermeasures}
    \centering
    \begin{tabular}{@{}c p{0.8\columnwidth}@{}}
        \toprule
        \textbf{\em Scale} & \textbf{\em Definition} \\
        \midrule
        0 & The notification does not mention countermeasures. \\
        1 & Countermeasures are incomplete or too advanced, e.g., \enquote{Browse the system log for indications of an attack.} \\
        2 & Unspecific but working countermeasures are described, e.g., \enquote{Disconnect the router from the network.} \\
        3 & Specific measures are explained step by step, e.g., \enquote{Unplug the router, perform a factory reset, and install a new firmware.} \\
        4 & Intuitive explanations of specific measures do not leave room for misunderstandings, e.g., describe in detail how to perform a factory reset and install an update on a certain router. \\
        \bottomrule
    \end{tabular}
\end{table}

For example, the user could simply turn off the threatened device.  
Much better countermeasures allow the user to eliminate a device's vulnerability, particularly if the countermeasure is intuitively explained step by step.

\paragraph{Correctness}

Dimension \textbf{Correctness} (see Table~\ref{tab:dim:correctness}) considers whether the dimensions of consequences, context, countermeasures, and urgency of the cybersecurity alert are neither missing, flawed, hallucinated, misleading, incorrect, nor described in a way that leaves room for mistakes for a user without cybersecurity expertise.

\begin{table}[h]
    \caption{Definition of the dimension \enquote{Correctness}.}
    \label{tab:dim:correctness}
    \centering
    \begin{tabular}{@{}c p{0.8\columnwidth}@{}}
        \toprule
        \textbf{\em Scale} & \textbf{\em Definition} \\
        \midrule
        0 & Consequences, context, countermeasures, or urgency are either missing, hallucinated, misleading, or incorrect, so that serious cybersecurity risks persist. \\
        1 & Consequences, context, countermeasures, or urgency are flawed or misleading, but this can be recognized with some research. \\
        2 & Incorrect or inconsistent consequences, context, countermeasures, or urgency can be recognized easily, e.g., if the notification mentions a device that is not in the network. \\
        3 & Consequences, context, countermeasures, and urgency are essentially correct, but the wording leaves room for mistakes. \\
        4 & Consequences, context, countermeasures, and urgency are correctly and unmistakably described. \\
        \bottomrule
    \end{tabular}
\end{table}

The rating of this dimension is based on the impact on cybersecurity. For example, a flawed countermeasure that has such an impact would be to stop warning messages about blocked network connections by disabling the router's firewall. On the other hand, an example of correct urgency is a notification that unmistakably explains how quickly a threat could result in which kind of harm to the home.

\paragraph{Intuitiveness}

Dimension \textbf{Intuitiveness} (see Table~\ref{tab:dim:intuitiveness}) measures whether the notification uses intuitive wording. This relates to the user's assumed lack of knowledge regarding cybersecurity-specific terms.

\begin{table}[h]
    \caption{Definition of the dimension \enquote{Intuitiveness}.}
    \label{tab:dim:intuitiveness}
    \centering
    \begin{tabular}{@{}c p{0.8\columnwidth}@{}}
        \toprule
        \textbf{\em Scale} & \textbf{\em Definition} \\
        \midrule
        0 & Consequences, context, countermeasures, or urgency are either missing or contain deep cybersecurity technical terms, e.g., \enquote{HTTP Response abnormally chunked.} \\
        1 & Some information related to consequences, context, countermeasures, or urgency is not intuitively understandable, e.g., \enquote{ntalkd might have a vulnerability hackers could exploit.} \\
        2 & Countermeasures and urgency are intuitively understandable, which allows the user to mitigate an attack without understanding it. \\
        3 & Context, countermeasures, and urgency are intuitively understandable, which allows the user to assess and mitigate the attack. \\
        4 & All parts of the rewritten notification are concise and understandable, without referring to deep cybersecurity terms. \\
        \bottomrule
    \end{tabular}
\end{table}

For example, we do not expect the user to be familiar with the names of attack vectors, specific threats, network protocols, Linux daemons, or network services. Intuitiveness and correctness meet at rating 0 (unsatisfactory), because missing information is unintuitive and incorrect at the same time. Our scale reflects that it is less of a problem if users don't understand the attack, as long as they can fix it properly.

\paragraph{Personalization}

Dimension \textbf{Personalization} (see Table~\ref{tab:dim:personalization}) considers to what extent the notification is personalized to the user, their use case, and home network.

\begin{table}[h]
    \caption{Definition of the dimension \enquote{Personalization}.}
    \label{tab:dim:personalization}
    \centering
    \begin{tabular}{@{}c p{0.8\columnwidth}@{}}
        \toprule
        \textbf{\em Scale} & \textbf{\em Definition} \\
        \midrule
        0 & The notification does not refer to the user or the network setup. \\
        1 & The notification is less specific and broad, e.g., \enquote{Anomalous actions are often first indicators of compromised devices.} \\
        2 & The notification is tailored to the user and their network, e.g., \enquote{The attacker could gain unauthorized access to your Echo Hub, potentially stealing sensitive information or using it to attack other networks.} \\
        3 & The notification is tailored to the user and their network and also refers to the specific mode of attack, e.g., \enquote{The malware Linux.IoTReaper tries to infect your Echo Hub, and could use it to attack others from your network.} \\
        4 & The notification includes comprehensive information about the user, the devices under attack, and the compromised use case, e.g., \enquote{Dear John, Linux.IoTReaper scans networks for vulnerable Linux devices and attempts to log into the devices. After that, the malware installs itself onto the system and begins downloading and executing commands from (...)} \\
        \bottomrule
    \end{tabular}
\end{table}

Thus, we assess whether a user can relate a cybersecurity threat to their actual situation. This refers to the network, its connected devices, and how the devices are configured and used. For example, assume a session-hijacking attempt on a smart security camera. By relating this alert to their concrete installation, the user can decide whether this is a threat to this specific camera or not. If the camera is disallowed from connecting to external devices anyway, the alert can be ignored.

\paragraph{Urgency}

Dimension \textbf{Urgency} (see Table~\ref{tab:dim:urgency}) determines how well the notification takes into account the urgency of dealing with the cybersecurity threat.

\begin{table}[h]
    \caption{Definition of the dimension \enquote{Urgency}.}
    \label{tab:dim:urgency}
    \centering
    \begin{tabular}{@{}c p{0.8\columnwidth}@{}}
        \toprule
        \textbf{\em Scale} & \textbf{\em Definition} \\
        \midrule
        0 & The notification does not address the urgency of action. \\
        1 & The urgency is communicated in unspecific, broad terms, e.g., \enquote{It is important to secure the network.} \\
        2 & A level of urgency is communicated, e.g., \enquote{The detected attack does not directly threaten your Echo Hub.} \\
        3 & Urgency is communicated and explained, e.g., \enquote{It’s important to take action quickly. Here’s why: (...)} \\
        4 & Urgency is communicated and explained, and also considered in the writing style of the countermeasures, e.g., \enquote{Your Echo Hub is under attack. It is important to quickly disconnect it from the network, before the attacker installs malware.} \\
        \bottomrule
    \end{tabular}
\end{table}

For example, ongoing attacks may require an immediate response, while an alert about a vulnerability that is not currently being exploited may allow for a certain delay. Outstanding (4) is a notification that not only tells the level of urgency but also uses wording for the entire message that reflects how quickly a response to the alert should be made.

\section{Case Study}
\label{sec:case}

In this section, we demonstrate HCSAEF's applicability for multifaceted analyses with three use cases:  
\textit{\enquote{Comparing Different Prompts}} for prompt optimization,  
\textit{\enquote{Comparing Different LLMs}} for explaining cybersecurity issues, and  
\textit{\enquote{Robustness of the Response}} of the LLM.

\subsection{Use Case: Comparing Different Prompts}

We exemplarily chose two different prompts, as shown in Table~\ref{tab:prompts}.  
\textbf{Prompt~1} follows a \textit{zero-shot} prompting approach, i.e., it consists of a single prompt without any examples of what the output should look like. The prompt is supplemented by a list of devices in the network that might be under attack.  
\textbf{Prompt~2} uses \textit{impersonation}, \textit{zero-shot chain-of-thought} prompting, and reflection~\cite{schulhoff2024prompt}, i.e., it tells the LLM to deliver an output according to a certain role, contains contextual information about the general setting, and urges the LLM to reason and reflect on its answer. However, Prompt~2 does not deliver details of the devices on the network to an externally hosted LLM.

\begin{table}[htb]
    \caption{Prompts}
    \label{tab:prompts}
    \centering
    \begin{tabular}{@{}c p{0.85\columnwidth}@{}}
        \toprule
        \textbf{\em No.} & \textbf{\em Prompt} \\
        \midrule
        1 & Explain the cybersecurity alert [ALERT] to a user without cybersecurity expertise. Align your explanation to the following home devices: Blink Surveillance IP Camera, Echo Hub, Smart TV, Game Console, Blink Camera, Shelly Motion Sensor. Provide countermeasures to isolate the attack and mitigate the possible damage. The countermeasures should be easy to follow and only involve tasks that are practicable for users without cybersecurity expertise. Tell the urgency of the issue. \\
        \midrule
        2 & Consider a private home network containing various smart home devices. A network monitoring tool has detected suspicious network traffic and generated the alert [ALERT]. You are a cybersecurity expert with years of professional experience in incident response. Your task is to write a message to the homeowner, which informs them about this alert. The homeowner has no prior knowledge in computer science and cybersecurity. Find out if the alert means that the network is currently under attack. Communicate the current threat to which the network is exposed. Explain which devices could be under attack. Explain how urgent it is to deal with the alert, and communicate the consequences of not reacting. Provide a detailed step-by-step plan of measures to mitigate the security threat. Reflect on your message and ensure that it is easy to follow for a user without technical expertise. \\
        \bottomrule
    \end{tabular}
\end{table}

To find out with HCSAEF which prompt is better suited for a user without cybersecurity expertise to deal with a potential cyberattack, we replace [ALERT] with the example from the introduction, \enquote{HTTP Response abnormal chunked for transfer-encoding}. Then, we execute both prompts with ChatGPT 4o~\cite{openai2024gpt4o} and evaluate its output with HCSAEF.  
Figure~\ref{fig:gpt4oresponse} exemplarily shows GPT 4o's output for Prompt~2 with the alert \enquote{\em HTTP Response abnormal chunked for transfer-encoding}.  
Table~\ref{tab:evalprompt2gpt4o} shows the result of this evaluation.

\begin{table}[htb]
    \caption{Evaluating Figure~\ref{fig:gpt4oresponse} with HCSAEF.}
    \label{tab:evalprompt2gpt4o}
    \centering
    \begin{tabular}{@{}l c p{0.5\columnwidth}@{}}
        \toprule
        \textbf{\em Dimension} & \textbf{\em Rating} & \textbf{\em Rationale} \\
        \midrule
        Consequences    & 3 & Consequences are specific and detailed to the extent of the information provided in the prompt. \\
        Context         & 3 & Context is specific but lacks some detail, e.g., what does \enquote{sending or receiving data in an abnormal way} mean? \\
        Countermeasures & 4 & Meaningful countermeasures are provided and explained. \\
        Correctness     & 4 & The rewritten alert carefully explains that abnormally chunked transfer encodings are not an attack as such, but might be an indication that an attacker is trying to find a weak spot on a device. \\
        Intuitiveness   & 4 & The rewritten alert only uses technical terms at an intuitive level. \\
        Personalization & 2 & Although no devices were mentioned in the prompt, the rewritten alert refers to typical devices that could be at risk. \\
        Urgency         & 4 & The rewritten alert explains in detail that an attack may be underway, which needs to be dealt with urgently. \\
        \bottomrule
    \end{tabular}
\end{table}

The table indicates that Prompt~2 indeed produces a notification that helps an everyday user secure their home network. However, there is room for improvement regarding the context of the attack, more specific consequences, and personalization. HCSAEF shows that it is worth considering providing the prompt with more details about the network and the user.

\begin{figure}[htb]
    \centering
    \includegraphics[trim=3.75cm 18.5cm 4.5cm 3.0cm, clip, width=0.95\columnwidth]{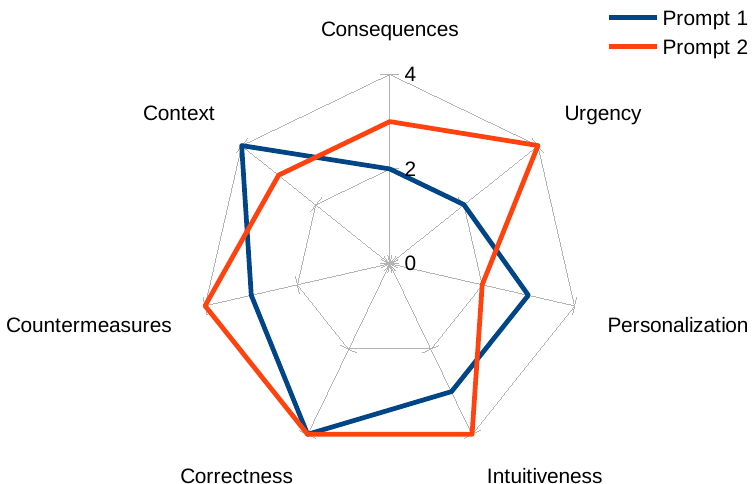}
    \caption{Comparing Prompt 1 and Prompt 2 with HCSAEF.}
    \label{fig:evalp1p2}
\end{figure}

Figure~\ref{fig:evalp1p2} compares the output of Prompt~1 and Prompt~2, both generated with GPT 4o.  
Prompt~1 uses a simpler prompting scheme than Prompt~2 but adds details about the network, as suggested by Table~\ref{tab:evalprompt2gpt4o}. For brevity, we refrain from reproducing the rewritten alert and the rationale for HCSAEF's assessment.

Figure~\ref{fig:evalp1p2} shows that adding further details indeed increases the ratings for Context and Personalization.  
However, with a simpler prompting scheme, the LLM produced a coarser output. For example, the LLM did not use the provided details about the devices to explain which cybersecurity risks exist due to the detected irregularities, and where to look for a reset button or firmware updates. With Prompt~1, however, the LLM generated a more general output and just mentioned the devices in an unspecific way. The countermeasures included tasks that require expertise, e.g., \enquote{\textit{Disable unused remote access features on your devices.}}, resulting in a lower rating for Intuitiveness.

We conclude that HCSAEF indeed provides a differentiated evaluation of security alerts rewritten by an LLM. This helps when tuning the prompts and deciding whether to provide details regarding installed devices and network configurations.

\subsection{Use Case: Comparing Different LLMs}

To evaluate how well each LLM explains a cybersecurity alert to everyday users, we ran experiments in March 2025 using the public web interfaces of the respective platforms. We tested Grok3 (grok-3-latest)\cite{xai2024grok}, GPT4o (chatgpt-4o-latest)\cite{openai2024gpt4o}, OpenAI o1 (o1-2024-12-17)\cite{openai2023gpto1}, and DeepSeekR1 (deepseek-r1:671b)\cite{deepseek2024} with Prompt1. All models were used with default settings, without fine-tuning or system modifications. Each received the same zero-shot prompt including the alert and network device details. For brevity, we summarize key output differences without reproducing full responses.

\begin{figure}[h]
    \centering
    \includegraphics[trim=3.25cm 18.5cm 5cm 3.0cm, clip, width=0.95\columnwidth]{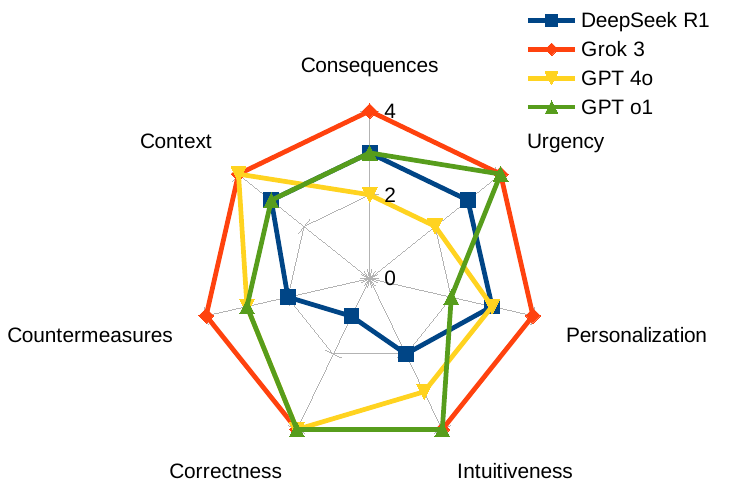}
    \caption{Comparing different LLMs with HCSAEF.}
    \label{fig:evalllms}
\end{figure}

Figure~\ref{fig:evalllms} shows the ratings of the LLMs we tested with Prompt1. Grok3 outperformed all other LLMs, using the devices in the prompt to explain in detail the attack consequences, how to narrow down infected devices, and how to perform a factory reset. It also conveyed the urgency clearly, stating, \enquote{\textit{This isn’t a drop everything and panic situation, but it’s serious enough to act on quickly—think of it like noticing a stranger hanging around your front door.}}

In contrast, DeepSeek~R1 generated misleading countermeasures that would provide new vulnerabilities, e.g., suggesting that the password for the security camera should be reset to \enquote{\textit{C@meraSunset2024}}, which an attacker could brute-force with a dictionary quickly. DeepSeek~R1 also delivered superficial and less complete consequences and assumed that any device in the network performs a factory reset by pressing the reset button for 10 seconds.

We already discussed the performance of GPT~4o in the last subsection. GPT~o1 performed slightly better. Its extended reasoning provided a more elaborate list of consequences of ignoring the alert. It also did not need technical terms to explain the cybersecurity threat and related countermeasures in precise language. However, GPT~o1 did not use the devices given in the prompt to generate a personalized answer. Instead, GPT~o1 restricted itself to general (but correct) explanations and countermeasures, such as \enquote{\textit{Keep an eye on your devices for unusual behavior—like random reboots, significantly slower performance, or new apps that you never installed on your Smart TV or Game Console. Weird changes often hint at malicious activity.}}

We conclude that HCSAEF generates a well-differentiated picture of the abilities of various LLMs to explain complex cybersecurity alerts. It seems that there are big differences in how the LLMs evaluate the same prompt, and selecting the proper model is an important step.

\subsection{Use Case: Robustness of the Response}
To find out how robust the generated responses are, we repeated Prompt~1 with Grok~3 and GPT 4o three times each. We did not modify the default \enquote{temperature} parameters. We observed that Grok's answers did not deviate much from one execution to another. Sometimes, the order of the countermeasures changed, and there were variations in the wording. Occasionally, Grok~3 decided to provide emotional support (e.g., \enquote{\textit{You don’t need to be a tech wizard to handle this!}}) or indicate the effort needed (e.g., \enquote{Check for Updates (,,,) Time: 10-15 minutes per device (plus update download time)}). All of Grok's responses were rated \enquote{Outstanding} in each dimension, with one exception: Once, Grok suggested a weak, dictionary-based password (\enquote{\textit{Set a new password (...) like MyDogRocks2025!}}).

\begin{figure}[h]
    \centering
    \includegraphics[trim=2cm 18cm 5.6cm 3.0cm, clip, width=0.95\columnwidth]{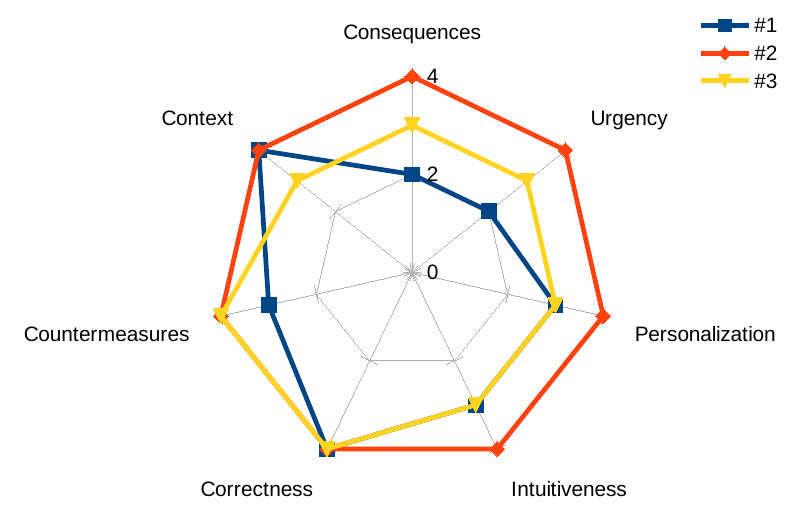}
    \caption{GPT 4o executing Prompt~1 three times.}
    \label{fig:evalmultiple}
\end{figure}

In contrast, GPT~4o's responses deviated significantly from one execution to another. It sometimes decided to consider the list of devices in the prompt and provided a personalized response, including a detailed step-by-step guide on how to execute a factory reset on each device named in the prompt. Since we executed our case study at different times of the day, we suspect that GPT~4o produces a more sophisticated response at times of lower system load. Figure~\ref{fig:evalmultiple} shows the evaluation of three executions of Prompt~1 with GPT-4o.

We conclude that HCSAEF allows us to observe important properties regarding the robustness of the prompt executions, which will foster fine-tuning the model or adjusting the temperature settings. For example, we observed GPT 4o generating heterogeneous responses, but all of them were correct.

\section{Conclusion and Future Work}
\label{sec:conclusion}

The proliferation of smart devices has made cybersecurity tools like firewalls and IDS relevant to everyday users. LLMs have been proposed to rewrite the technical alerts of security tools into actionable notifications that are intended to help private users secure their homes. This work introduces HCSAEF, which allows for the evaluation of such notifications across seven dimensions. The purpose of HCSAEF is to support multifaceted analyses, such as comparing the capabilities of different LLMs in explaining cybersecurity issues, different prompting strategies, or whether providing more details to the LLM actually leads to better notifications. We have demonstrated HCSAEF's applicability through a case study.

For the time being, we have evaluated HCSAEF's dimensions manually. Our next step will be implementing HCSAEF into a RAG approach, i.e., we will generate a synthetic evaluation data set as a reference and use an LLM-as-a-judge approach to automatically evaluate cybersecurity notifications. Once automated, we will use HCSAEF for large-scale experiments with various rewriting approaches, prompting strategies, and LLMs. Furthermore, we plan to run comparative experiments to determine whether HCSAEF's evaluation is similar to the assessment of a human user, in order to fine-tune the rating and build a ground truth for future evaluations.

\section*{Acknowledgment}
\addcontentsline{toc}{section}{Acknowledgment}
We sincerely thank Louis Carlos Roth for his invaluable assistance with the evaluation framework and the case studies. 

\bibliographystyle{IEEEtran}
\bibliography{publications-reduced}

\end{document}